\documentclass[namedreferences]{solarphysics}
\usepackage[optionalrh,solaromanenum,natbib]{spr-sola-addons} 
\usepackage{graphicx}                    
\usepackage{color}                       
\usepackage{url}                         
\usepackage{fancyvrb}

\newcommand{\apj}{    {\it Astrophys. J.}}
\newcommand{\nat}{    {\it Nature}}
\newcommand{\grl}{    {\it Geophys. Res. Lett.}}
\newcommand{\aap}{    {\it Astron. Astrophys.}}
\newcommand{\jgr}{    {\it J. Geophys. Res.}}
\newcommand{\solphys}{{\it Solar Phys.}}
\newcommand{\ssr}{    {\it Space Sci. Rev.}}
\newcommand {\etal}{{\it et al.}}
\newcommand {\arcsec}{{\hbox{$^{\prime\prime}$}}}
\newcommand {\arcmin}{{\hbox{$^{\prime}$}}}
\newcommand*\Del{\mathrm{\Delta}}

\begin{document}

\begin{article}

\begin{opening}

\title{The SUMER Data in the SOHO Archive}

%
\author{W.~\surname{Curdt}$^{1}$\sep
        D.~\surname{Germerott}$^{1}$\sep
        K.~\surname{Wilhelm}$^{1}$\sep
        U.~\surname{Sch\"uhle}$^{1}$\sep
        L.~\surname{Teriaca}$^{1}$\sep
        D.~\surname{Innes}$^{1}$\sep
        K.~\surname{Bocchialini}$^{2}$\sep
        P.~\surname{Lemaire}$^{2}$
       }

\runningauthor{Curdt \etal}
\runningtitle{The SUMER data in the SOHO archive}

%
  \institute{$^{1}$ Max Planck Institute for Solar System Research (MPS)\\
                     email: {\sf curdt@mps.mpg.de}\\
             $^{2}$ Institut d'Astrophysique Spatiale (IAS)\\
                     email: {\sf karine.bocchialini@ias.u-psud.fr} \\
             }


\begin{abstract}

We have released an archive of all observational data of the VUV spectrometer
{\it Solar Ultraviolet Measurements of Emitted Radiation} (SUMER) on SOHO that
has been acquired until now. The operational phase started
with `first light' observations on 27 January 1996 and will end in 2014.
Future data will be added to the archive when they become available.
The archive consists of a set of raw data
(Level~0) and a set of data that are processed and calibrated to the best knowledge
we have today (Level~1). This communication describes step by step the data
acquisition and processing that has been applied in an automated manner to
build the archive. It summarizes the expertise and insights into the scientific
use of SUMER spectra that has accumulated over the years. It also indicates
possibilities for further enhancement of the data quality. With this article
we intend to convey our own understanding of the instrument performance to the
scientific community and to introduce the new, standard-FITS-format database.
\end{abstract}


\end{opening}

\section{Introduction}
\label{sec: Intro}

The {\it Solar Ultraviolet Measurements of Emitted Radiation} (SUMER; Wilhelm~\etal, 1995)
telescope and spectrometer has been taking ultraviolet spectra (50~nm to 161~nm) from
the {\it Solar and Heliospheric Observatory} (SOHO) since its launch in December 1995.
The complete dataset up to January 2013 has been reprocessed and is now available in FITS format
from the SOHO archive.
The motivation and basic idea behind this communication is to describe aspects
of the data acquisition that are relevant to the data quality and
details of the various steps and procedures needed to arrive at Level~1
(L1) data. The data reduction steps used in the archive correspond to the best
knowledge of today. Even after so many years, the processing is still not perfect
and at several stages compromises had to be taken, which may explain why this
task was not completed earlier.

\begin{figure}
\centerline{\includegraphics[width=1\textwidth,clip=]{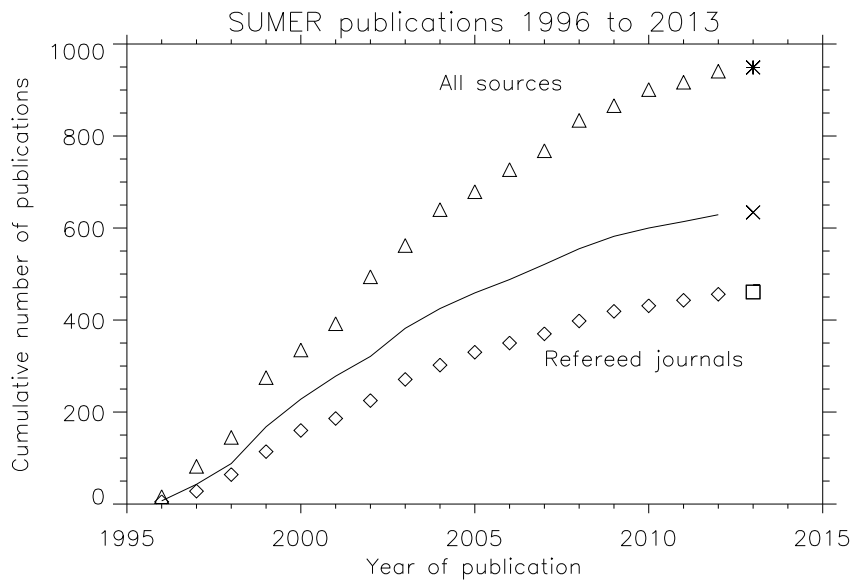}}
\caption{Record of refereed publications. Entries listed by ESA under SUMER/SOHO
are given as solid line, whereas those in ADS are denoted as diamonds.
The year 2013 is shown with special symbols, because the figures are not final.
The graph stands for the experience accumulated over the years, which is the base of the archive.}
\label{fig:publ}
\end{figure}

SUMER data has so far been accessible from archives that contain reformatted
telemetry, {\it i.e.} raw, Level~0 (LZ) data that is provided either in FITS, FTS,
or IDL restore format. Various correction and calibration procedures have to
be applied in order to minimize the known instrumental effects in LZ data and
to convert the incoming signal to physical quantities of the radiation.
The basic knowledge acquired during ground testing and commissioning was
continuously extended, confirmed, or improved over the years. This knowledge
is documented in almost 1000 articles in refereed journals as indicated in
Figure~\ref{fig:publ}. There are clear indications from the publication rate
that also future work with SUMER data can be expected. It is our motivation
to make sure that the accumulated knowledge will not be forgotten, and at the
same time provide ready-to-use spectra to the community for future work.

Details of the reduction procedures were made publicly available on the instrument web sites.
The latest update was in 2008.\footnote
{http://www.mps.mpg.de/projects/soho/sumer/text/cookbook.html}
The procedures used to produce L1 data described here
represent the present view of our instrument. A fair amount of this
information is presented as a complete, overworked and self-standing document.
We cannot exclude deviations from the results of data processing that was
completed in earlier times. The differences are to our knowledge small and will in
all likelihood not affect the validity of previous analyses. Therefore, this improved
view of our instrument in no way compromises reduction work with previous
versions, but does also not claim to be final.
For this reason we have tried to make the description of each individual step as
transparent as possible. This may help the user not to use the reduction as a black box,
but to improve the results if future insights allow him to do so.

The application of the various correction and calibration procedures has to
reverse the order applied to the incoming signal by the various
instrument subsystems. This applies in particular to stages of the signal processing
in the detectors with specific shortcomings (procedures 2 to 6).
Therefore the sequence must be executed in the described order:
\vspace{2mm}
\newline
    1. Decompression\\
    2. Deadtime correction\\
    3. Odd-even pattern\\
    4. Local-gain correction\\
    5. Flatfield correction\\
    6. Geometric distortion correction\\
    7. Radiometric calibration
 \vspace{2mm}
 \newline

If at a later time, an improvement in any one of these procedures is
possible, then either the raw data can be reprocessed with the improved
algorithm or the existing correction must be re-tracked to the problematic
procedure that has to be replaced and re-run together with those following
in the sequence. All information needed for the re-tracking is
documented in the FITS header of L1 data.

In addition to these procedures that affect the pixel values themselves, we
also mention procedures that are to be used for special analyses, but have not
been applied to the archive data. These are procedures to
compensate for the instrumental effect on the width of spectral lines, to
compute the movement of the slit image as parasitic effect of the grating
focus mechanism, to estimate the stray-light level in off-limb spectra,
and give information on the wavelength calibration including thermo-elastic
deformation effects along the spectral dimension.

\section{Instrument Description}
\label{sec: Instrument}

The instrument is described in \citet{Wilhelm95}. Performance details
are given in \citet{Wilhelm97a} and \citet{Lemaire97}.
The concept of the instrument is also discussed and reviewed in the general context of
VUV space instrumentation in \citet{Wilhelm04}. A survey of the most
significant results is found in \citet{Wilhelm07}. Here we emphasize those
details that are relevant to the archive.
The particular spectral range that is covered by SUMER comprises emission
lines and continua of many elements. It includes the entire Lyman series of
hydrogen and chromospheric emission from other atomic constituents, as well as
emission lines useful for observations of transition-region or coronal phenomena.
It turned out that forbidden transitions between high-lying levels of
highly ionized iron and other heavy species can be used as proxies for X-ray
radiation and tracers of processes during flare events \citep{Feldman00},
thus supporting X-ray spectroscopy.

\subsection{Optical Design}
\label{sec: Optics}

\begin{figure}
\centerline{\includegraphics[width=0.7\textwidth,clip=]{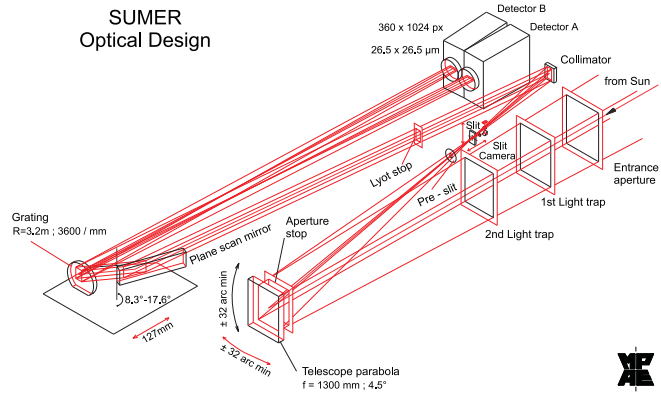}}
\caption{Layout of the optical system. The diagram shows the extreme rays of the folded beam
and indicates the position and movement of the mechanisms serving the system.}
\label{fig:opt}
\end{figure}

The optical design of the instrument is shown in Figure~\ref{fig:opt} and
discussed in \citet{Wilhelm95}. Many features hereof are relevant for
instrumental effects and are important for the data reduction. The optical
system is based on a normal-incidence off-axis telescope and a slit spectrometer
in Wadsworth mount. Pointing is accomplished by two nested mechanisms that allow,
for optical performance reasons, a spherical motion of the parabolic mirror
around the changeable slit in the focal plane. The beam issued from the
slit -- entrance of the spectrometer -- is collimated by an off-axis parabolic mirror.
The collimated beam is seen by a spherical concave grating producing a stigmatic
image of the slit that is dispersed in wavelength. A plane mirror in front of
the grating allows us to change the angle of incidence and thus, to modify the
setting of the instantaneous wavelength portion seen by the detectors in the
focal plane of the grating. The effective focal length of the grating depends
on the angle of incidence and thus on the wavelength setting. A grating focus
mechanism is, therefore, needed to keep the spectral image on the detector always in focus.
For this reason, important optical parameters are wavelength dependent,
notably the magnification, the angular pixel size along the slit,
and the dispersion defining the spectral pixel size. All these
parameters are provided in the FITS header of the data files.

The reduction of scattered light has been a strong requirement for the
optical system and the surface quality of the optical components with
chemical-vapor-deposited (CVD) silicon carbide (SiC) coated surfaces.
Measurements of the surface roughness performed at
GSFC (Saha and Leviton, 1993) have shown that the rms micro-roughness at 10 $\mu$m
scaling length was 0.6~nm, which is well within the specification.
Such measurements have been used to predict the scatter performance at FUV wavelengths.

\subsection{Mechanisms}
\label{sec: Mechanisms}

SUMER is equipped with seven mechanisms, the door mechanism, two mechanisms
for pointing in azimuth and elevation, the slit changer, the telescope focus mechanism,
the wavelength scan, and the grating focus.
All mechanisms are actuated by stepper motors and have position encoders for
monitoring purposes. The encoders are operated in open-loop modes, {\it i.e.}, the
positions that are read out are telemetered as housekeeping values, but not
used to control the stepper motors.

\subsubsection{Pointing in Azimuth and Elevation}
\label{pointing}

With incremental single steps of {0.38\arcsec} both in azimuth and elevation,
the telescope could be pointed anywhere in the field of view of
{64\arcmin}~$\times$~{64\arcmin} centred on the Sun.
The azimuth drive was also used to scan regions of interest on the Sun and to
compensate for the solar rotation in sit-and-stare applications. The nominal
elevation pointing position is always related to the central pixel of the slit,
irrespective of the location of the slit image on the detector.

The absolute pointing uncertainty of typically {10\arcsec} is the combined result of
thermoelastic effects on SOHO as well as in the instrument, of step losses in
the mechanism, and of the parasitic shift of the slit image on the detector
resulting from a misalignment of the grating focus mechanism (as described below).
From time to time and for special cases the solar limb was used to `re-calibrate'
the reference position of the SUMER pointing mechanism
(the zero position of the SUMER coordinate system).

When step losses of the azimuth drive occurred in October 1996, it was decided
to operate the driving motor with retaining power and in high-current mode.
Only sporadic step losses occurred over the succeeding years in this mode.
However, in April 2008, the problem re-appeared ex nihilo and became worse since then.
Raster scans could not reliably be completed anymore. With the help of the
azimuth position encoder, pointing was still possible. The value
of the azimuth and elevation encoders are given in the L1 image header in order to
improve the knowledge of the pointing. We note, however, that the encoder reading for
the housekeeping process is completed at a cadence of 15~s and not
synchronized with the science operation. With short exposure times it may happen
that the encoder values in the first image header are not yet updated.

\subsubsection{Slit Focus and Slit Select}
\label{slit}

During the commissioning phase in early 1996, the slit focus mechanism was used to
optimize the focus position of the telescope \citep[{\it cf.},][]{Lemaire97}.
The setting has not been changed since then. The slit changer can choose between
the four slits of size {4\arcsec}~$\times$~{300\arcsec},
{1\arcsec}~$\times$~{300\arcsec}, {1\arcsec}~$\times$~{120\arcsec}, and
{0.3\arcsec}~$\times$~{120\arcsec}.
Images of the short slits can be positioned in such a way that the top, the central,
or the bottom section of the detector active area are illuminated, `top' or `bottom'
settings are called `asymmetric' slit positions. In bottom position, a baffle obscures the
extreme pixels of the short slits (\#5 and \#8). The slit selection constrains
the image format that is appropriate for the detector readout. Possible image formats
including their telemetry load are listed in Table~\ref{tab:formats}. The low telemetry
rate of $10.5 \times 10^3$~bits per second turned out to be a major limitation for
fast data acquisition. Faster raster scans could, however, be completed by
buffering spectra into the on-board memory up to a data volume of $\approx 5 \times 10^6$~bytes.

\begin{table}
\caption{Standard image formats. Other formats that are only used for special applications
are  not listed here. The telemetry-load column applies for the nominal TM rate
of $10.5 \times 10^3$~bits per second.}
\label{tab:formats}

\begin{tabular}{rrrlr}
\hline\noalign{\smallskip}
ID & Spectral pixels & Spatial pixels & Format & TM load\\
&&&& s\hspace{5mm} \\
\noalign{\smallskip}\hline\noalign{\smallskip}
2 & 1024 & 360 & 1 byte  & 280.9 \\
3 & 1024 & 360 & 2 bytes & 561.8 \\
4 & 1024 & 120 & 1 byte  & 93.7 \\
5 & 1024 & 120 & 2 bytes & 187.3 \\
8 & 50 & 360 & 1 byte    & 13.8 \\
9 & 50 & 360 & 2 bytes   & 27.5 \\
10 & 50 & 120 & 1 byte   & 4.6 \\
11 & 50 & 120 & 2 bytes  & 9.2 \\
12 & 25 & 120 & 1 byte   & 6.9 \\
13 & 25 & 120 & 2 bytes  & 13.8 \\
14 & 25 & 120 & 1 byte   & 2.3 \\
15 & 25 & 120 & 2 bytes  & 4.6 \\
37 & 256 & 360 & 2 bytes & 140.5 \\
38 & 512 & 360 & 1 byte  & 140.5 \\
39 & 512 & 360 & 2 bytes & 280.9 \\

\noalign{\smallskip}\hline
\end{tabular}
\end{table}

\subsubsection{Wavelength and Grating Focus}
\label{grating}

\begin{figure}
\centerline{\includegraphics[width=0.7\textwidth,clip=]{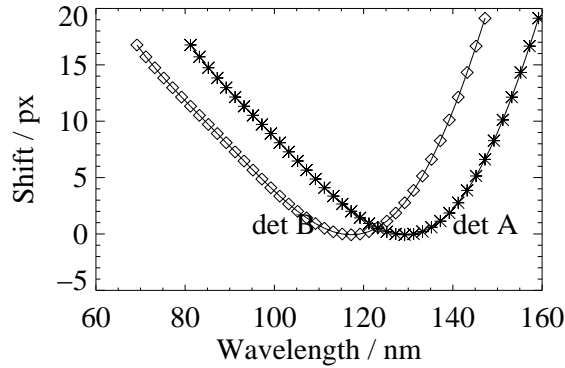}}
\caption{Parasitic pixel displacement of the slit image over the spectral range.}
\label{fig:shift}
\end{figure}

The optical system requires that the wavelength and grating focus mechanisms are
always operated simultaneously. Because of a misalignment of the grating focus
drive -- a combination of the misalignment of the guiding rails of the grating focus mechanism,
the rotational axis of the wavelength mechanism, and the grating optical axis --
the position of the slit image on the detector is slightly offset, whenever a new
wavelength setting is commanded. The combined effect of this parasitic movement
and the change of the angular pixel size is up to 25~pixels as shown in
Figure~\ref{fig:shift} (see also Section~5.4). Since the readout window is fixed and
does not follow the shift, often dark pixels appear in short-slit image formats.

When only partial spectral windows are transmitted, the line of interest
should in an ideal case be centred in the telemetered window. This is normally
not the case and one should keep in mind that the repeatability of this mechanism is
limited (see Section~5.4.1). At short wavelengths, one actuator step corresponds to
seven spectral pixels, while several steps are needed for a shift of one
pixel at the other extreme. Therefore, problems with the wavelength
setting occur more often in the short wavelength range.

\subsection{Detectors}
\label{sec: XDL}

SUMER is equipped with two photon-counting detector systems. Details of the
detectors and their performance are given in \citet{Siegmund94}.
A triple stack of multichannel plates (MCP) carries the photocathode deposited on
the front face of the first MCP and a cross delay-line anode converts photons to
an electronic pulse. The travel time of the pulse through the crossed delay-lines
determines the location of the photon event and the image is constructed by a
time-to-digital converter (TDC) that creates a 1024~$\times$~360 array of photon events,
which we for simplification, but inaccurately call pixels. It is this
analog-to-digital conversion and, in particular, the linearity of this ADC which,
in part, is responsible for some of the artifacts in the digital image,
most notably the odd-even pattern, as described below.

High overall count rates result in deadtime effects, since every individual
event has to be processed by the post-anode digital electronic and bright lines
will lead to local gain depression of the MCPs. The adverse effects of these shortcomings
are balanced or even overcompensated by a unique feature of photon counting systems:
their dark signal is almost negligible so that deep exposures can be made
with very low signal. In other words, the dynamic range can be extended
over many orders of magnitude.

\begin{figure}
\centerline{\includegraphics[width=1\textwidth,clip=]{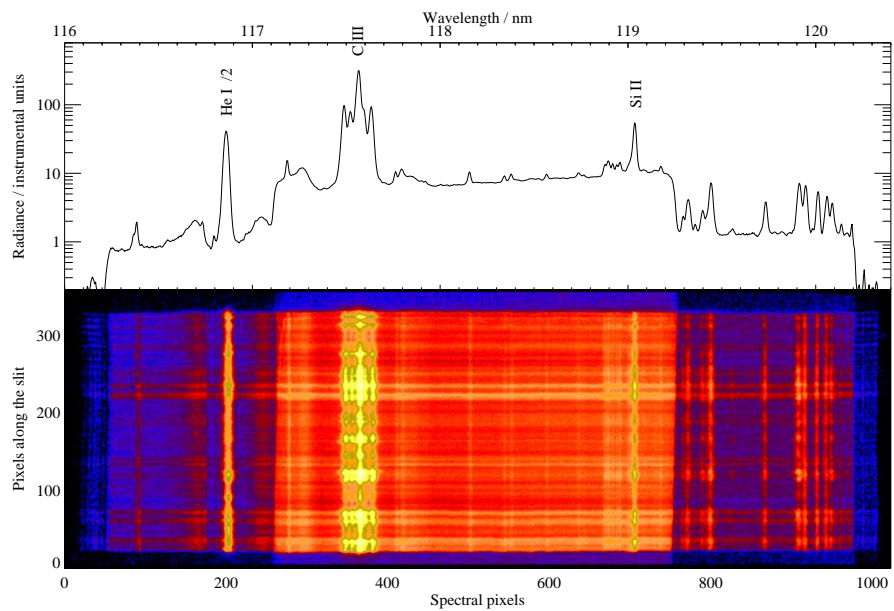}}
\caption{Sample data of a full frame image showing the detector layout. The image
has been reversed in wavelength and detector effects have been corrected.
The spectral range from 116.0~nm to 120.4~nm comprises the He\,{\sc i} line at 58.4~nm,
observed in second order of diffraction.
In this spectral range, the efficiency of the central, KBr-coated section is significantly enhanced.
The 1:10 attenuation affects 50 pixels on both extremes
({\it cf.}, Section~5.4.1 for wavelength calibration). The image is part of a QS
dataset obtained on 20 April 1997 \citep[{\it cf.},][for details]{Curdt01}.}
\label{fig:xdl}
\end{figure}

A sample spectrum around 118~nm is shown in Figure~\ref{fig:xdl} to illustrate the
layout of the detector array. Only the most prominent lines are indicated
\citep[{\it cf.},][for a comprehensive line identification]{Curdt01}.
The slit image covers only $\approx$300 out of the 360 spatial pixels.
The central spectral pixels ($\approx$280 to $\approx$770) represent those sections,
where the KBr photocathode is deposited on the bare MCP. These pixels have a much higher sensitivity,
in particular in the spectral range from 90~nm to 130~nm (see Section~6 for
details). Some pixels at the bare-to-KBr transition are difficult to interpret,
since this is not a sharp boundary. The extreme 50~pixels on both sides are
covered by a grid that serves as an 1:10 attenuator. However, as a side effect the attenuation
exerts a modulation on the line profile, which makes it difficult to interpret these data.

SUMER is also equipped with a rear slit camera (RSC). Despite misalignment problems,
which reduced the scientific value of this device, the RSC could be used to verify
the pointing mechanism by locating sunspots and by observing the solar limb.
In the archive, RSC images are only included as raw data.

\subsection{Instrument Operation}
\label{sec: Operation}

The principal data acquisition features of SUMER are described in detail in
\citet{Wilhelm95}.
The instrument operation applied during the many years of observation is summarized hereafter.
During several years (1996 to 2003), the 24~h of SOHO throughput operation was divided in
a long pass with continuous real-time access during at least 8~h and two or
three small 1~h real-time passes. Later, the real-time coverage was reduced and
became more irregular. The round-trip time for real-time commands is about
10~s with an uplink rate of three bytes per second. The on-board SUMER memory
capability was 64 elementary or macro commands (formatted as User Defined Programme,
see the following UDP section). So, at any time during a real-time pass we have been able to
load instantaneous or time-tagged activities. The primary ground station to
operate the instrument is located at the GSFC/EOF (Experiment Operation Facility
at the Goddard Space Flight Center). During few campaigns remote terminals at the
IAS/MEDOC (Multi-Experiment Data and Operation Centre at the Institut d'Astrophysique Spatiale)
could be connected to this station. Since 2011 we can also operate our instrument remotely from MPS.
The operations were completed following a timeline comprising a time span of
$\approx$24~h. The definition of the timeline was accomplished in several steps.

\subsubsection{Target Selection and Definition of the Type of Observation}
Target selection was done by looking at the status of the Sun with the image-tool
routine. Image-tool is the user interface of a pointing tool that is based on a database
of solar images obtained from ground-based and space observatories,
{\it e.g.} SOHO/{Extreme ultraviolet Imafing Telescope} (EIT; Delaboudini\`ere
\etal, 1995) and allows fine targeting for an observation to come hours or days in
advance.\footnote{http://hesperia.gsfc.nasa.gov/ssw/gen/idl/synoptic/image\_tool/image\_tool.pro}
The choice of the feature (coordinates and times) to be observed
was done by the SUMER observer or in coordination with other SOHO (or other
ground-based or space observatory) instruments.

\subsubsection{SUMER Command Language, Predefined and User-Defined Programmes}
A dedicated command language has been used to define the structure and contents of
observing sequences. A library of high-level functions specific to SUMER and elements to
build loops, branching points, etc. are the basic features of the SUMER command
language (SCL). Various mapping modes and other elements of the SCL library are described in
\citet{Wilhelm95}. The SCL source code to define an observing sequences
is written in plain English. It sets the telescope pointing coordinates, the selected entrance slit of
the spectrometer, the reference pixel on the detector associated with a wavelength,
the wavelength and the associated windows on the detector, the exposure times,
the number of spectra to be collected, and the detector voltage handling.
A set of 30 differently structured, hard-wired observing sequences, so-called
Predefined Observational Programmes (POPs) was already included in the SUMER flight
software. Some had a simple linear structure, but more complex ones
with loops and branching points were also available.
In addition, new User-Defined Programmes (UDPs), also written in SCL code,
could be added to the available POPs after passing different stages of a
validation process. First, the syntax was checked, then the code was
compiled and converted to a token code as input to the TKI (Token Code Interpreter)
programme, a tool that is also part of the on-board software. Finally, the token code
of each new UDP had to pass an instrument simulator before the sequence was
added to the pool of more than 1000 validated UDPs. In addition to the POPs,
the token code of 16 different UDPs could be held on board at the same time.

Unfortunately, a flaw in the detector communication occasionally terminated
the full execution of a POP. Mitigation of this problem required to modify the
hard-wired structure of the code and this could only be completed by rewriting
the sequence as UDP. The availability of such user-defined sequences demonstrated
the enormous flexibility of the software design and turned out to be highly
appropriate for the operation of SUMER. Most of the time, SUMER was operated
in this mode. Any spectrum in the archive is linked to its `mother' UDP,
which is kept in the UDP database ({\it cf.}, Section~2.6).

\subsubsection{Generation and Uploading of Commands}
The prepared observation programmes are inserted into the SUMER Planning Tool,
which can generate a file of time-tagged execution commands to be sent to the
SUMER on-board computer through a SOHO channel opened by the SOHO/EOF ground system.
At the same time, an activity plan can be produced to be loaded into the
SOHO activity data base, documenting the plan to be executed.

\subsubsection{Monitoring and Preprocessing of Real-Time Data}
A few seconds after the observation the real-time
data are collected and displayed on ground computers (EOF, MEDOC, MPS).
Few minutes after the data are taken, they are pre-processed using a
quick-look facility to follow the progress of the observations. That has been
very useful in order to react in near real-time to the selection of sunspot
coordinates, the solar limb position (using the rear slit camera) or the selection
of any solar feature within a raster scan image. In parallel house keeping data
are received in real-time and are used to follow the instrument parameter and
to detect any anomaly that requires an operator reaction,
either to check the accomplishment of the programme and re-iterate the observation
as needed or to optimize parameters for a secondary run of that programme.

\subsection{Ground Support Facilities}
\label{sec: EGSE}

The ground support facilities were built as a dual Electrical Ground Support
Equipment (EGSE): a scientific EGSE based on a VMS operating system
(here referenced as operation station) and a PC-based maintenance EGSE.
The maintenance EGSE is used for real-time commanding and real-time visualization
of scientific and housekeeping data necessary to check the health
of the instrument and the running of the observational programme. The operation
station was used to complete scientific observations: target selection, creation
of UDP, insertion of individual commands and programmes in the time line through the Planning Tool,
sending the commands, reception of the telemetry, pre-processing the raw scientific data in
the Quick-Look and formatting the zero level FITS (FTS) data to feed the
preliminary SOHO archive. The dual EGSE was duplicated at the IAS/MEDOC centre.

\subsection{UDP Database}
\label{sec: UDP}

The source code of all used UDPs is accessible on the MPS SUMER archive web page.
These files are not needed directly for the FITS formatting process. Information
about, which programme is used to acquire the dataset is included in the
FITS image header (see Section~\ref{sec: Aux}).

\subsection{The Archive}
\label{sec: Archive intro}

The SUMER archive is part of the total SOHO archive which is built around
the archive created by all SOHO instruments.
The main SOHO archive is based at NASA/GSFC, while a copy is maintained at ESA
and at IAS/MEDOC. The access to the SUMER LZ archive can be accomplished through
various web pages.\footnote{
{\tt sohowww.nascom.nasa.gov/data/archive, sohowww.estec.esa.nl/data/archive/, idc-medoc.ias.u-psud.fr}}

\section{Data Resources}
\label{sec: Data}

In this chapter it is described which data sources are included in the processing
from TM raw data to the ready FITS product. Further on it describes the various
additional data sources which are processed and how this information is added to the
FITS image header. A flow diagram showing the data sources and processing
steps is depicted in Figure~\ref{fig:dataflow}.

\subsection{TM Processing}
\label{sec: TM processing}
The SOHO telemetry data is distributed in files which contain the data of one day
(see SOHO Interface Control Document\footnote{\tt sohowww.nascom.nasa.
gov/publications/soho-documents/ICD/icd.pdf}).
For SUMER there are three different kinds of file, standard housekeeping data
called HK0 and science low and high rate files. The HK0 data are extracted
out of the TM files and written, also on a daily base, into
binary files. This is more or less a copy process. Processing the science data
needs considerably more effort, since the SUMER image data packets are interlaced with
science housekeeping packets ({\it e.g.} HK255, see SUMER Operations
Guide\footnote{\tt www.mps.mpg.de/projects/soho/sumer/text/sum\_opguide.html},
Chapter~5 for more detailed information). During processing the various packets are extracted and,
depending on the type, are sorted into binary files for HK and images.
The produced binary HK0, science HK, and image files build the base for all further
processing and SUMER data products.

\subsection{Science Data}
\label{sec: Science data}


The main sources of SUMER data are the SOHO/SUMER telemetry (TM) files.
These files consist of a header, the TM packets, information about telemetry
gaps and information about the transmission quality of the packet as listed below.
The gap and quality information are added to the SUMER image header in the flags
QAC and MDU. If the QAC flag is set it means that during transmission an error
occurred. When the MDU flag is set it means, there are TM packets missing and
the image contains fill data at these positions. The fill data is chosen as the
max value in the image. For traceability of the data processing, the name of
the TM source file is added to the FITS header.
\newline
The reflecting FITS keywords are:
\vspace{3mm}

\begin{tabular}{l l}
Keyword&Description\\ \hline
XSSMDU&Missing data in image 0=no,1=yes \\
XSSQAC&Quality of image data 0=OK,1=NOTOK \\
XSSFID&File ID from TM file catalog \\
XSSFPTR&Pointer to image position in bin file \\
XSCDID&CD ID of TM file \\
XSSEQID&CD sequence ID of TM file \\
XSTMFILE&TM filename without ext \\
\end{tabular}

\subsection{Housekeeping Data}
\label{sec: HK}
The SUMER image header already contains a snapshot of most housekeeping values
of the moment the image is taken. But there are still significant values missing, like
temperatures of the instrument and the encoder positions of the mechanisms.
These values are sampled every 15~s to 45~s and are sent asynchronously via the housekeeping
channel HK0. The HK0 data are read and correlated to the start time of an exposure
plus half of the exposure time so that they are accurate mid-way through the
exposure.  This approach is more accurate for exposure times longer than
1~min because of the sample rate of HK0 data. The maximum delay for the HK0 values is 300~s.
This information is added to the FITS Header.

\vspace{3mm}

\begin{tabular}{l l}
      Keyword&Description\\ \hline
      T3TELE&Telescope (MC2) temperature in degree Celsius \\
      T3REAR&SUMER rear (MC3) temperature in degree Celsius\\
      T3FRONT&SUMER front (MC4) temperature in degree Celsius\\
      T3SPACER&SUMER spacer (MC6) temperature in degree Celsius\\
      MC2ENC&SUMER MC2 (azimuth) encoder position\\
      MC3ENC&SUMER MC3 (elevation) position\\
      MC4ENC&SUMER MC4 (slit select) position\\
      MC6ENC&SUMER MC6 (scan) encoder position\\
      MC8ENC&SUMER MC8 (grating) encoder position\\
      HK0TIME&Time stamp of HK0 record\\
\end{tabular}
\clearpage

\subsection{Auxiliary Data}
\label{sec: Aux}

The information about UDP name, campaign, etc. are extracted from the
Oracle planning database and commanding log files. This data is formatted into FITS files (tables).
This intermediate extraction step is done due to the fact that the SUMER
planning database is not globally accessible, and the produced FITS files
can be more easily distributed.
These files will be put into the SSWDB area so the the information can be accessed
with simple FITS read programmes ({\it e.g.} {\bf fits\_read}). The task of
correlating the information by time, is done during processing (LZ preparation).

The routine {\bf update\_fits} completes the time correlation and adds the
information to the FITS header. This routine calls the routines
{\bf get\_fitsudpinfo} and {\bf get\_fitscmpinfo}, which read the information
from the auxiliary FITS tables.

\vspace{4mm}

The information/keywords added to the header by {\bf update\_fits} are:\\


\begin{tabular}{l l}
Keyword&Description\\ \hline
CMP\_NAME&Name of campaign observation \\
CMP\_ID&Campaign number \\
STUDY\_ID&Study number (database ID) \\
STUDY\_NM&Study name \\
OBJECT&Target \\
SCIENTIST&Scientist responsible of POP/UDP \\
OBS\_PROG&Name of scientist programme \\
PROG\_NM&Name of observing programme \\
\end{tabular}

\vspace{4mm}

The spacecraft attitude information are distributed as FITS files.
Using the IDL routines {\bf get\_sc\_att} and {\bf get\_sc\_point} from Solar Software tree (SSW)
the attitude information can be read and then added into the FITS header.
The solar angles P0 and B0 are read with the routine {\bf pb0r} and also added
to the header.
\vspace{.5cm}

\begin{tabular}{l l}
      Keyword&Description\\ \hline
SOLAR\_P0&Solar angle P0 / deg \\
SOLAR\_B0&Solar angle B0 / deg \\
\end{tabular}
\vspace{4mm}

Geocentric and heliocentric information is added from information
read in with the function {\bf get\_orbit}.


\clearpage
\section{Data Reduction Steps}
\label{sec: Reduction}

Details of the data reduction steps that were used to prepare L1 data are
described here. The order of the various data reduction steps follows the
rational as explained before.

\subsection{Decompression}
\label{sec: decompression}

In most cases, the accumulated counts of the 16-bit data array have been
compressed on board to 1-byte integers. The by far most often applied
`Method~5' (quasi-logarithmic byte scale)
used an algorithm that mapped the dynamic range found in the image
to a logarithmic lookup table. For numbers from 0 to 107 the result of the lookup
table is a one-to-one copy of the input value.  Thus all values with low
count rates are preserved losslessly, and can be
reversed by a decompression that uses information contained in the raw image header.
Data in LZ format as well as in FITS and FTS format of existing archives are
already decompressed during creation, if methods between 5 and 10 were employed.
Data that was compressed by application of more
complex compression methods ({\it e.g.} `moment calculation') is only available in
LZ format, since decompression is not possible.

\subsection{Reversion}
\label{sec: Reversion}

The pixel addresses of the SUMER detectors are such that the highest
wavelength is on pixel 0, the lowest on pixel 1023; the wavelength stated
in the raw header refers to the reference pixel. Therefore, wavelengths
are descending from left to right in reformatted raw data.
For compatibility with the SOHO conventions and the following image correction routines,
the spectral direction of the images in LZ format of this archive and in FITS or FTS
format has already been reversed (such that wavelength increases to the right).

\subsection{Flatfield Correction and Removal of Digitization Nonlinearity}
\label{sec: FF}


\subsubsection{Image Features that Need a Correction by a Flatfield Routine}

Detector anode and photocathode both show effects that are specific for each
detector and are relevant for the data reduction. The flatfield correction
of SUMER images generally corrects small-scale structures introduced by the
detectors. Larger structures, like the overall response of the different
photocathode areas, are treated by the radiometric calibration routine.
The small-scale structures are introduced by the spatial inhomogeneity of the channel plate
response and the non-linearity of the analog-to-digital converters of the detector
electronics. Another small-scale structure that may be present in SUMER images,
is local gain depression which, however, must be treated by another routine.

Inhomogeneity of the micro-channel plate response is mainly due to the hexagonal
pattern of the microfiber bundles and the relative orientation of these bundles
in a stack of three microchannel plates. Depending on this relative orientation,
a complex moir\'e pattern of the response is produced, which is very distinct in
the detector~A and much less pronounced in the detector-B. If a clear hexagonal
(`chicken wire') pattern is visible in the image it is produced by the lowest
channel plate (the one which is closest to the anode). The smaller structures
are produced by the superposition of the fiber bundles of the three plates.

In addition to the moir\'e pattern there may be dead pores in one of the channel plates,
which lead to dark spots in the image. Depending on which of the plates has the
dead pore determines the size of the dark spot. Because of the spreading of the charge,
when transferred from one plate to the next, more pores are blacked-out in the
succeeding plate and, thus the dark spot is larger when it is in the first plate
(the one farthest from the anode).

The further processing of the signal to create a digital image out of the analog
signal by the time-to-digital converter (TDC) leads to additional small-scale
artifacts in the image data. A non-linearity of the analog-to-digital-converter (ADC)
introduces a difference in the response between two succeeding rows of the image.
This can be seen in the very distinctive alternating response of the odd and even
rows of the image. The ADC non-linearity causes that the signal in one row is
about 9.5\,\% higher than average, while in the adjacent row it is 9.5\,\% lower than average,
thus making on average a 19\,\% difference between the rows. In the following,
we will call this the `odd-even pattern'.
Note that this pattern only exists in the rows of the image, not in the columns
where it has been avoided internally by the detector electronics using an
unsharpening technique (`dithering'). This effect, which is present in both detectors,
is also effectively averaged out if an even number of binning is applied along
the slit direction.
The non-linearity of the time-to-digital converter of the post-anode digital
electronics is found to be constant in time and well characterized for both detectors.
It can rather easily be removed. However, during the period from June 2005
to November 2006, when a flaw in the address decoder of detector~A developed
that rapidly progressed and finally led to the destruction of the TDC unit,
the fixed pattern of this chain increased significantly.
During this time period, the characteristic of the TDC was regularly monitored,
so that these data are still scientifically sound. The best results were found
by separation of both effects, when in a first step the pattern is compensated
and in a second step the residual non-uniformities of the pixel array were removed.

\subsubsection{Producing Flatfield Correction Data}

There is no flatfield illumination of SUMER detectors on board. There are
however alternative ways in producing quasi-flat illuminations of the SUMER
detectors that can be used to extract the small-scale features and produce a
data array that can be used to compensate these artifacts.

\begin{figure}
\centerline{\includegraphics[width=1\textwidth,clip=]{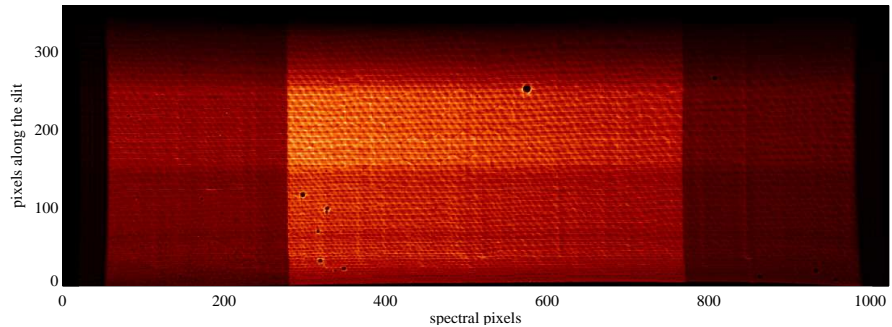}}
\caption{Flatfield exposure of the detector~A showing the moir\'e pattern
of the MCPs and some smaller dark features. The dark spot is a cluster of low-gain pores.
Note, extreme pixel addresses around the edges and the areas of the attenuators are not useable.}
\label{fig:ff}
\end{figure}

In order to produce a quasi-flat illumination of the detectors on board, an
observing sequence has been written that puts the SUMER spectrograph in a state
of maximal defocusing at the wavelength of 88~nm for a several-hour long
observation of a preselected quiet-Sun (QS) area at the wavelength of the Lyman
continuum. At this wavelength the solar spectrum is devoid of strong lines,
and a quiet solar region avoids bright features with strong variability in the
field of view. The defocusing provides a smearing of any small features to a
size of at least 16 pixels. Thus, the resulting image of this deep exposure
-- an example is shown in Figure~\ref{fig:ff} --
can be used to extract small-scale features of the detectors. An on-board routine
extracts from this exposure the small-scale `fixed pattern' by applying a median
filter (of size 16~pixels) to the data and division by the filtered image. The
resulting image (the flatfield array) is a matrix with values between 0.5 and 1.5,
the average value being 1.0. It is stored on board for processing of images and
sent to ground by telemetry for further application to any data on the ground.
The flatfield correction array is applied in a simple multiplicative way.
The procedure of flatfield correction can be made on ground (or, reversely,
the on-board flatfield corrections can be undone) by using the function
{\bf sum\_flatfield.pro} available in the Solar Software Tree.

Flatfield arrays have been produced frequently until 1998, but later the occasions
of making flatfield exposures have been greatly reduced in an effort to reduce
the total exposure of the detectors to high count rates that would lead to
ever increasing high voltage needed to compensate for the resulting gain loss.
As the high voltage power supply units were reaching their upper limits, at
some occasions only short flatfield exposures have been taken, which could be
used to find any changes in the flatfield structures with respect to the latest
deep exposures.

Another method to determine the `fixed pattern' of the detectors can be
employed when a large enough data set can be used to extract from the average
of all images, in a similar way as above, the small-scale features introduced by the detector.
In general, this average of the images is not as deep an exposure as the `normal'
flatfield exposure of three hours in the Lyman continuum, but its one certain advantage
is that it is as close in time as possible to the actual observations to which it
will be applied. This is useful in particular cases, because not all of the
`fixed patterns' are really fixed in time.

\subsubsection{Changes of the Fixed Pattern with Time}

It had been found very early in 1996, by comparing several flatfield correction
arrays of detector~A, that the flatfield pattern of the detectors were changing
slightly with the time of usage of the detector. This can be found by correlating
with each other different flatfield array data. The correlation can be maximized
by a shift of the flatfield pattern, which is mostly less than one pixel
(in $X$- and $Y$-directions), but can amount to several pixels between different flatfields.
There may be several reasons for this shift of the small-scale pattern: One reason
lies in the `scrubbing' of channel plates, i. e., extracting charge from the MCPs
by heavy usage, and the resulting gain loss of the lowest of the three channel plates.
Since the channels are inclined with respect to the anode, the gain loss may
cause a shift of the charge cloud centroid that should be located on the anode
by the position encoding. Such an effect was clearly detected when a strong
spectral line was observed for long time at the same location on the detector.
Then, the persisting gain loss at this location resulted in a shift of the charge
cloud centroid towards the area with higher gain. Fortunately, the SUMER wavelength
scan mechanism allows the position of spectral lines to be place anywhere on the
detector and this resulted on the average to a more or less uniform gain loss
across the active area. When such a uniform gain loss was reached, increasing
the high voltage of the MCPs could compensate it. This gain calibration has
been done frequently.
But the change of high voltage may have an effect on the electrical field
between the channel plate and the anode. It may also have an effect on the
position encoding if it affects the travel time of pulses in the cross-delay
lines of the position encoding system. Both may lead to a shift of the image pattern.
Therefore, new flatfield images have been acquired regularly - roughly every month -
after the high-voltage setting had been newly adjusted during a gain calibration.
Later, when the observations of the solar disk have been reduced to save
lifetime of the detectors, the period between flatfield acquisitions has been increased.

Since the gain loss occurs more or less constantly during usage of the detector
and the compensation can only be done stepwise, there is, in principle, a possible
small shift between the data and the flatfield pattern. But in general the shift
of the flatfield pattern is not uniform. A uniform scrubbing of the detector
cannot be achieved, and therefore a differential (or local) scrubbing, which
is due to the non-uniform illumination of the detector during its use, results
in a shift pattern that is not uniform: depending on which part of the detector
area has been used more, the shift is higher in these areas.
In addition, as mentioned above, the `fixed pattern' results from a superposition
of the pattern of each channel plate. The scrubbing, however, takes place mostly
in the lowest channel plate (the one closest to the anode), from which the largest
amount of charge has been drawn. Thus it may be possible that features arising from
different channel plates may suffer a different displacement in the image. However,
this intricacy may be difficult to detect, since the differences are probably much
smaller than one pixel.

There is, however, a very strong fixed pattern in the flatfield data that never changes.
It is the nonlinearity of the detector ADC in the position encoding electronics,
which causes the difference of responsivity of odd and even rows. This odd-even
pattern is always present along the slit direction, and it has been found to be
very stable throughout the time of all flatfield images we have.

\subsubsection{Application of the Flatfield Correction}

The general flatfield routine {\bf sum\_flatfield.pro} applies to the image data the
flatfield correction array in the same way as the on-board flatfield routine.
It corrects all fixed pattern by multiplication of the flatfield correction array.
It does not take into account any changes of the fixed pattern with time.
Thus, it corrects perfectly the odd-even pattern and much of the other channel
plate non-uniformities. By selecting the flatfield array dated closest to the
date of observation, the shift between the flatfield data and the corrected data
can be minimized. This is the simplest approach, and for most purposes the
results are satisfactory.

To improve the flatfield correction, we can take the shift of the fixed pattern
into account. In this case, the odd-even pattern must be corrected first.
For this purpose we have extracted
the odd-even pattern from the flatfield raw data and produced new flatfield arrays
that have the odd-even pattern removed. This was done in the following way:
Separately for detectors~A and B, the average odd-even pattern was determined
from the row-sums of all flatfield exposure raw images available. From the row-sums,
the odd-even pattern was extracted by subtraction of the two-pixel average.
Since the pattern is a non-linearity of the ADC, it must be the same all along the slit.
Thus, the average along the row-sum was taken to determine a single value for the
upper and lower deviation, respectively, from the average. These two values were
taken to construct an artificial image array of the odd-even pattern of 1024 by 360 pixels.
This array can thereafter be used to remove the odd-even pattern from images by
multiplication (in the same way as the usual flatfield function). It has also been
applied to all the flatfield raw images, in order to remove from them the odd-even
pattern and to produce the new flatfield correction arrays without odd-even pattern.
In order to apply the shifted flatfield correction to SUMER data, there are now
odd-even arrays and flatfield arrays available to apply these corrections sequentially
(see the SUMER Data Cookbook for details about how to use these files).

\subsection{Geometric Distortion Correction}
\label{sec: Geo}

\begin{figure}
\centerline{\includegraphics[width=1\textwidth,clip=]{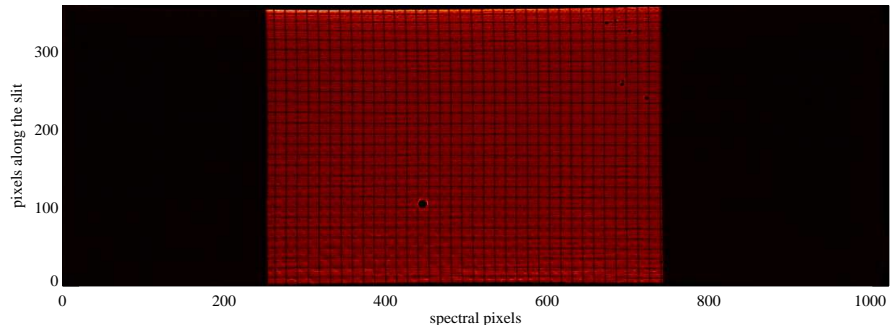}}
\caption{Image of a grid shadow placed in close proximity of the detector
during laboratory testing showing the cushion shape distortion. Note that the bright
central part is caused by the photocathode coating.  Dark pores are visible.
The grid structure has been used to determine the deviation of the image from
the ideal rectangular structure with a uniform plate scale.}
\label{fig:grid}
\end{figure}

The digital image created by the detector is not a perfectly rectangular array but,
due to the analog image acquisition method using pulse travel times and
time-to-digital conversion, it is distorted in a cushion shape, resulting in a
non-linear spatial and spectral scale. Most of this distortion is due to
inhomogeneity in the anode causing small differences of the propagation speed
of pulses in the delay lines. This leads to local variations of the plate scale
of the detector (Wilkinson~\etal, 2001). The distortion correction for both
detectors is based on images of a rectangular grid that was placed in front of
the detectors before integration into the SUMER instrument. Figure~\ref{fig:grid}
shows one of these images that have been used to determine the correction matrices.

In addition to the geometric distortion, the spectral lines are inclined with
respect to the detector vertical lines due to a discrepancy in the orientation
of the grating and the detector. For precise wavelength measurements a highly
accurate linearization of the spectral plate scale is necessary. Thus, the image
distortions need a geometrical correction such that the curvature of spectral
lines is removed and the wavelength scale is made linear. A combination of the
laboratory images and data from cool solar emission lines acquired for this
particular purpose have been used to extract the needed information for creating
the geometric correction arrays (Moran, 2002).
The arrays consist of lookup tables with pixel shift vectors that can be
applied by using a bilinear interpolation algorithm to correct the distorted
frames with a standard uncertainty of 0.11~pixel in spectral and 0.25~pixel
in spatial direction.
This algorithm incorporates resizing of the pixels while maintaining the radiometric
accuracy. As a side effect of this treatment, empty values will be produced in
some pixels near the edges. This led to an adverse effect in those cases
where adjacent windows were selected, since these empty pixels may cut out
important parts of the line of interest. Therefore we had to concatenate
individually read out windows to produce full-size formats. These `inserted'
full image formats led to a significant blow up of the data volume of the archive.

The distortion correction for both detectors is based on images of a
rectangular grid that was placed in front of the detector before integration.
This may explain, why a residual distortion still remains for edge pixels.

\subsection{Dead-Time and Local-Gain Effects}

The \emph{total} count rate of a detector during one exposure may be so high
that individual photon events may not be detected correctly and electronic
deadtime correction factors must be applied. The deadtime
effect of the detector electronics is not negligible whenever the total count
rate on the detector is above 50\,000~s$^{-1}$. The {\bf deadtime-corr.pro}
routine takes care of this effect and corrects the radiometric calibration
using the total count rate as input. Note that the total count rate on
the detector cannot be inferred from subformat images.
Instead, this information is taken from the detector housekeeping channel.
Due to the cyclic readout of detector housekeeping data -- a process that is
asynchronous to the science observation -- the actual incoming
event rate may not be updated fast enough in the header data when a change of
photon flux happened less a minute before the exposure.

The \emph{local} count rate in a spectral line may also be high, such that the
local gain in this part of the detector channel plate is reduced and pulses may fall
below the detection threshold. This reduction in responsivity can be corrected
by the local-gain depression correction as long as the incoming photon flux
is moderate and the pixel count rate is below $\approx$20~s$^{-1}$.
For higher pixel count rates, the uncertainty increases dramatically. In case
of severe overexposure, the number of valid counts will be reduced instead of
going up. Although the scientific use of such spectra is questionable, they
appear unflagged in the archive.

\section{Radiometric Calibration}
\label{sec: Radio}

In this section, the radiometric calibration of the spectrometer SUMER and
related aspects will be outlined. The spectroradiometry to be employed is
covered in many publications that will be summarized here with reference to
the most relevant original articles.

The solar electromagnetic radiation\,---\,of which SUMER can observe the
wavelength range from 46.5~nm to 161.0~nm\,---\,can be characterized by the
total solar irradiance (TSI) and its spectral
distribution, the solar spectral irradiance (SSI), as a function of
the wavelength, $\lambda$ (or, alternatively, the frequency, $\nu$).
Quantitative information on these quantities can only be obtained with
calibrated instrumentation, {\it i.e.}
the observations must be compared to laboratory-based standards,
thereby providing a baseline for short-term and long-term investigations
of any solar variability ({\it cf.} Quinn and Fr\"ohlich, 1999; Lean, 2000;
Willson and Mordvinov, 2003; Wilhelm, 2009, 2010). The physical quantities have
to be given in units of the International System of Units
(SI: Le syst\`eme international d'unit\'es) (BIPM, 2006; see also NIST, 2008).

\subsection{Calibration Concept}
\label{concept}

In Table~\ref{Tab_Units}, some derived SI units of physical quantities are
compiled that are relevant in the context of spectroradiometry.

\begin{table*}
\caption{Derived SI units used in spectroradiometry.}
\begin{small}
\begin{tabular}{llll}
\hline
Quantity & Symbol$^{\rm a}$  & Unit symbol & Unit name of quantities\\
\hline
Radiant energy & $Q$ & J &
joule (1~J=1~kg\,m$^2$\,s$^{-2}$)\\
Radiant flux, power & $\mathit \Phi$ & W  &
watt (1~W=1~J\,s$^{-1}$)\\
Spectral flux & ${\mathit \Phi}_\lambda$ & W\,nm$^{-1}$ & \\
Irradiance & $E$ & W\,m$^{-2}$ &\\
Spectral irradiance & $E_\lambda$ & W\,m$^{-2}$\,nm$^{-1}$ &\\
Radiance & $L$ & W\,m$^{-2}$\,sr$^{-1}$ & \\
Spectral radiance & $L_\lambda$ & W\,m$^{-2}$\,sr$^{-1}$\,nm$^{-1}$ &\\
Radiant intensity & $I$ & W\,sr$^{-1}$ &\\
Spectral intensity & $I_\lambda$ & W\,sr$^{-1}$\,nm$^{-1}$ &\\
\hline
\end{tabular}

\vspace{0.2cm}
$^{\rm a}$ Recommendations only.
\end{small}
\label{Tab_Units}
\end{table*}
Four quantities are of major importance for the spectroradiometry:
the ``radiant flux density'' or ``irradiance'', $E$; the ``spectral
irradiance'', $E_\lambda$; the ``radiance'', $L$; and the
``spectral radiance'', $L_\lambda$. They are given in Table~\ref{Tab_Units}
together with supporting explanations. Note that the radiance and the
intensity are not dependent on the observing distance, whereas the irradiance
varies with the inverse square of the distance. The spatially-resolved
observations of SUMER yield the spectral
radiance, $L_{\lambda}(\vartheta,t)$, defined by the relation
\begin{equation}
{\rm d} Q(\lambda,\vartheta,t) =
L_{\lambda}(\vartheta,t)\,{\rm cos}\vartheta\,{\rm d}S\,
{\rm d}\omega\,{\rm d}t\,{\rm d}\lambda ~,
\label{radiance}
\end{equation}
where d$Q$ is the differential radiant energy emitted
into the solid angle ${\rm d}\omega$ from $\cos \vartheta\,{\rm d} S$,
the projected area normal to the direction of ${\rm d}\omega$,
during the time interval ($t,~t + {\rm d}t$)
in the wavelength interval ($\lambda, ~\lambda + {\rm d}\lambda$).
An average value of the spectral radiance, $\overline{L_{\lambda}}$, over
certain solid angle, time, and wavelength intervals
can be obtained from a measurement of the energy
\begin{equation}
\Del Q~=~\overline{L_\lambda}\,\Del \Omega\,\Del t\,\Del \lambda\,A
\label{energy}
\end{equation}
through the aperture
area, $A$, of SUMER ({\it cf.} Wilhelm, 2002a).
If the wavelength interval $\Delta \lambda$ covers
the profile of a spectral line at $\lambda$,
$L_{\rm line} = (\overline{L_\lambda} - L_{\rm back})\,\Delta \lambda$
represents\,---\,after a suitable background subtraction\,---\,its line
radiance.

As mentioned before, the radiometric calibration must be traceable to
laboratory standards. Ideally these would be primary standards\,---\,
absolute radiation sources that can be realized in the laboratory
(Smith and Huber, 2002). Synchrotron radiation constitutes a suitable
source standard in the VUV range, because the spectral radiant flux emitted
can be calculated from the parameters of the electron or positron storage
ring (Schwinger, 1949; Hollandt \etal, 2002).
In most cases, secondary standards have to be engaged
as transfer standards between primary standards and the instrumentation to
be calibrated, because the operational requirements of the primary
standard and those of the test specimen are often not compatible.

\subsection{SUMER Radiometric Calibration}
\label{radiometry}

A calibration of the SUMER spectrometer designed for operation on a spacecraft
directly at a synchrotron facility would have caused conflicts in
cleanliness requirements and schedule constraints.
A transfer standard equipped with a hollow-cath\-ode
plasma-discharge lamp was therefore calibrated with BESSY~I
at the PTB laboratory\footnote{Berlin electron storage ring for
synchrotron radiation; \newline Physikalisch-Technische Bundesanstalt}
for 32 emission lines with wavelengths between 53.7~nm and 147.0~nm.
This was done by a comparison of the radiation characteristics of both
standards with the help of a VUV monochromator, taking into account
the polarization of the synchrotron beam. The calibration of the transfer
standard was carried out before (and after) it was used to characterize
the spectral response of SUMER.
During the calibration runs, a reproducibility of the radiant flux could be
obtained in certain spectral lines within $\pm\,2.5\,\%$ over several hours,
and $\pm\,5\,\%$ after a change of the filling gas
(Hollandt \etal, 1996, 1998, 2002; Wilhelm \etal, 2000).

The laboratory calibration was intended to measure the radiometric
response of the system, as far as mirror reflectivities and detector
responsivities were concerned, without internal vignetting.
Contributions to the relative standard uncertainty by the various subsystems
have been compiled in Table~\ref{Uncertainties} for the central wavelength range.
The data resulted in relative standard uncertainties of 0.11 using the 2~mm hole
in place of the slit, 0.12 with the nominal slit and 0.18 for the
0.3\arcsec $\times$ 120\arcsec~slit. Based on these measurements
Figures~\ref{fig:cal}(a) and \ref{fig:cal}(b) have been plotted.
\begin{table*}
\caption{Contributions to the relative uncertainties
(wavelength range 53.7~nm to 123.6~nm).}
\begin{flushleft}
\begin {small}
\begin{tabular}{lll}
\noalign{\smallskip}
\hline
\noalign{\smallskip}
Item & Quantity & Uncertainty \\
     & &(1 $\sigma$) \\
\noalign{\smallskip}
\hline
\noalign{\smallskip}
Transfer standard & ($6.80 \times 10^6$ to $7.04 \times
10^8$) s$^{-1}$ & 0.06 to 0.07 \\
\hspace*{0.2cm} (photon flux) &&\\
Detector and & 5.64 mm$^2$/(9.5 mm $\times$ 27.0 mm)$^{\rm a}$  &\\
\hspace*{0.2cm} telescope inhomogeneities
& 140 mm$^2$/(90 mm $\times$ 130 mm)$^{\rm a}$ & 0.08 \\
Aperture stop & 90 mm $\times$ 130 mm & 0.001/0.001 \\
Focal length of telescope & 1302.77 mm at 75 $^\circ$C
& $5 \times 10^{-5}$ \\
Slits:  &&\\
\hspace*{0.2cm} \#1 (4\arcsec $\times$ 300\arcsec)
& 26.03 $\mu$m $\times$ 1889.7 $\mu$m & 0.005/0.003 \\
\hspace*{0.2cm} \#2 (1\arcsec $\times$ 300\arcsec; nominal)
& \hspace*{0.08cm} 6.23 $\mu$m $\times$ 1889.7 $\mu$m & 0.016/0.003 \\
\hspace*{0.2cm} \#4 (1\arcsec $\times$ 120\arcsec)$^{\rm b}$
& \hspace*{0.08cm} 6.27 $\mu$m $\times$ 755.4 $\mu$m & 0.016/0.005 \\
\hspace*{0.2cm} \#7 (0.3\arcsec $\times$ 120\arcsec)$^{\rm b}$
& \hspace*{0.08cm} 1.76 $\mu$m $\times$ 755.4 $\mu$m & 0.045/0.005 \\
\hspace*{0.2cm} \#9 ($\oslash$: 317\arcsec; calibration) &
2 mm diameter hole & 0.003\\
Nominal/calibration slit & 0.0110 (signal ratio) & 0.05 \\
Lyot stop &  27.73 mm $\times$ 40.12 mm & 0.004/0.003 \\
Slit diffraction & Model calculations & 0.01 \\
Detector pixel size (mean)
& 26.5 $\mu$m (spat.)$\times$ 26.5 $\mu$m (spectr.)
& 0.02/0.015 \\
Detector & Flatfield and distortion corrections & 0.01 \\
\hline
\multicolumn{3}{l}{\tiny a. Illuminated area during calibration at any time divided by
total size of optically effective area.} \\
\multicolumn{3}{l}{\tiny b. Slits \#3 (6) and 5 (8) are identical with slit \#4 (7),
but offset in spatial direction. The extreme}\\[-2mm]
\multicolumn{3}{l}{\tiny \hspace{3mm}pixels of slits \#5 and \#8 are vignetted.}
\end{tabular}
\end{small}
\end{flushleft}
\label{Uncertainties}
\end{table*}

The scattered-light measurements were carried out using a source emitting two
intense Kr\,{\sc i} lines at 116.5~nm and 123.6~nm, because they have
wavelengths close to the bright H\,{\sc i} Ly-$\alpha$ line at 121.6~nm.
The laboratory results showed excellent stray-light characteristics.
Nevertheless, the scattered light of the H\,{\sc i} Lyman-$\alpha$ and $\beta$
lines could be observed at 1700\arcsec~from the centre of the solar disk in
order to obtain line profiles unaffected by the geocorona (Lemaire \etal,
1998, Lemaire 2002; Emerich \etal, 2005).

\subsection{Calibration Tracking}
\label{tracking}

A critical issue is the stability of the spectroscopic responsivity,
which can be affected by obstructions of the apertures or contamination of
the optical elements and detectors. Solar ultraviolet radiation leads to
photo-activated polymerization of contaminating hydrocarbons and, as a result,
to a permanent degradation of the system. A cleanliness programme is therefore
of great importance to ensure particulate and molecular cleanliness of the
instruments and the spacecraft (Sch\"uhle, 1993, 2003; Thomas, 2002).
An instrument door and an electrostatic solar wind deflector
in front of the primary telescope mirror are specific features incorporated
in the design in order to maintain the radiometric responsivity during launch
and the operational phases.

\begin{figure}[th]
\includegraphics[width=12cm]{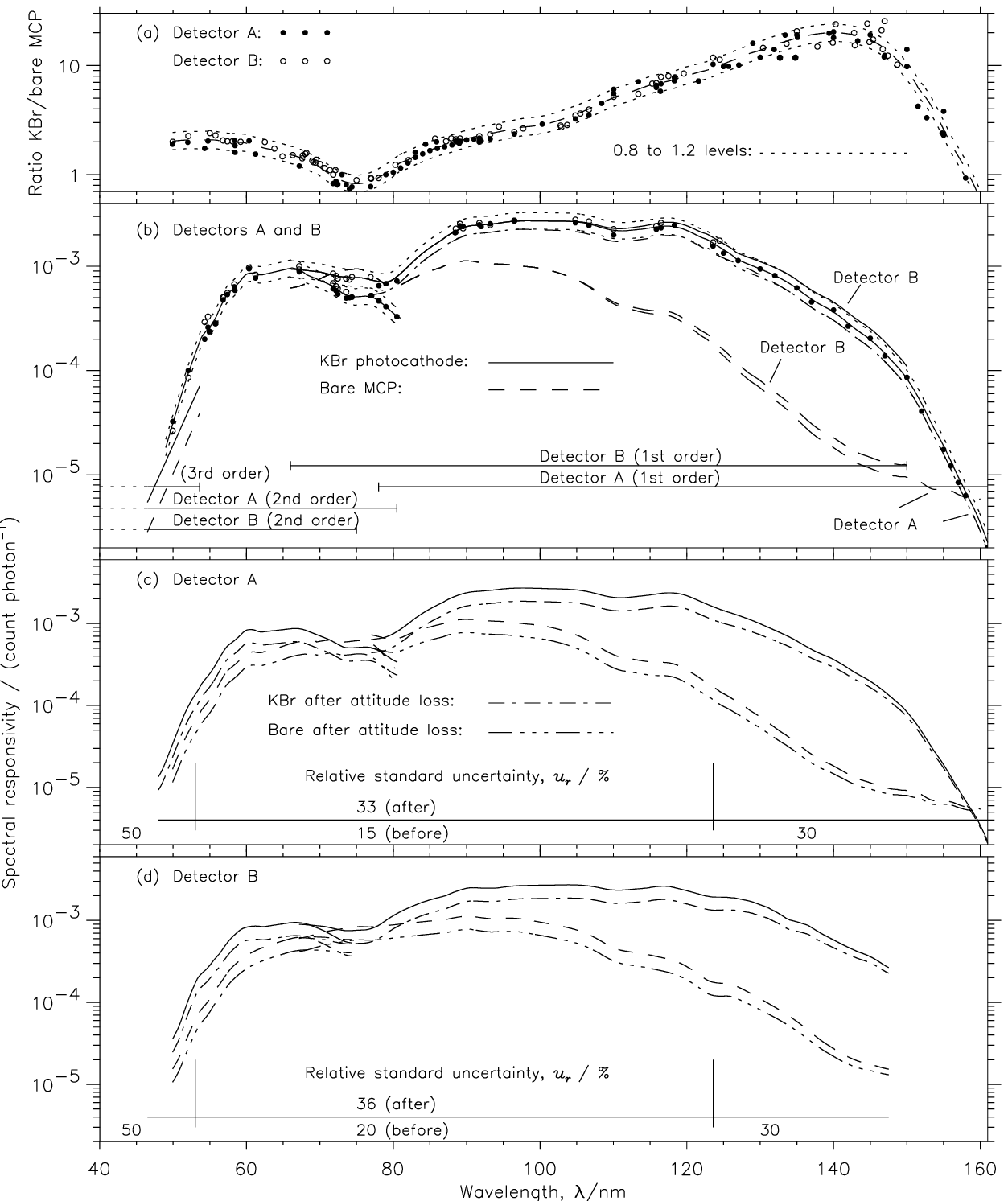}
\caption{\small
Spectral responsivities of the SUMER instrument with its detectors~A and B,
and the corresponding relative uncertainties for the nominal slit.
(a) In the upper panel the responsivity ratio of the photocathodes is shown.
(b) First-order and second-order KBr responsivities evaluated jointly for
both detectors. The long-wavelength deviation of detector~B is treated
in the text as well as the bare MCP responsivities and the third-order calibration.
Independent  assessments of (c) detector~A  and (d) detector~B.
For both detectors the relative uncertainties inside and outside the
wavelength band from 53.7~nm to 123.6~nm and their changes after the recovery of SOHO are indicated.
(On the long-wavelength side, the relative uncertainty refers to the KBr
photocathode only.)
}
\label{fig:cal}
\end{figure}

Nevertheless it is essential to track the calibration status through all
phases of the mission, {\it i.e.}, transport, spacecraft integration and tests,
launch, commissioning as well as operations.
The procedures include in-flight calibration using line ratios provided in atomic physics data
(Doschek \etal, 1999; Landi \etal, 2002),
inter-calibration between instruments (Wilhelm, 2002b), degradation monitoring
(Wilhelm \etal, 1997; Sch\"uhle \etal, 1998, 2000a):

\begin{enumerate}

\item In order to obtain a radiometric characterization outside the spectral
range covered in the laboratory, line-radiance ratios measured on the solar
disk have been compared with the results of atomic physics calculations.

\item A deep-exposure reference spectrum obtained with detector~A in a stable
coronal streamer on 13 and 14 June 2000 showed the Si\,{\sc xii} pair at
49.94~nm and 52.07~nm in second and third order of diffraction. This allows us
to establish responsivity curves in third order for both photocathodes.

\item Integration of the spectral radiance using full-disk SUMER scans
have been performed
in the N\,{\sc v} line at 123.8~nm and the C\,{\sc iv} line at
154.8~nm in 1996. The spectral irradiance of the Sun so obtained
could be compared with the
{\it Solar Stellar Irradiance Comparison Experiment} on the {\it Upper Atmosphere Research
Satellite} (SOLSTICE/UARS; Rottman \etal, 1993; Woods \etal, 1993)
radiometrically calibrated at the Synchrotron Ultraviolet Radiation Facility
(SURF-II) at NIST. Agreement within a factor of 1.14 was found for the
N\,{\sc v} line and approximately 1.10 for the
C\,{\sc iv} line (Wilhelm \etal, 1999).

\item Stellar observations (Lemaire, 2002) and reference spectra of QS
regions taken on both detectors indicated that the data points available for
both detectors are not systematically different within the relative
uncertainty margin of $\pm 20$\,\%. It was thus possible to determine joint KBr
responsivity functions for detectors~A and B in Figure~\ref{fig:cal}(b).

\item A stable radiometric calibration is also supported by the results of the
flatfield exposures performed in the H\,{\sc i}
Lyman continuum near 88~nm in QS areas. No significant change of the
responsivity of detector~A has been found over 200~days, nor was there any
decrease in detector~B over 350~days (Sch\"uhle \etal, 1998).

\item  However, a major change of the responsivity happened during the attitude loss
of SOHO in 1998. After the recovery, a responsivity decrease was found over a wide spectral
range as shown in Figures~\ref{fig:cal}(c) and \ref{fig:cal}(d).
This change is attributed to the deposition of contaminants and subsequent polymerization
on the optical surfaces of the instrument, because both detectors were equally affected.
Relative losses of 26\,\% for He\,{\sc i} 58.4~nm, 28\,\% for Mg\,{\sc x} (60.9 and 62.4)~nm,
34\,\% for Ne\,{\sc viii} 77.0~nm, 39\,\% for N\,{\sc v} 123.8~nm, and 29\,\%
for the H\,{\sc i} Lyman continuum were obtained, resulting in an average relative loss
of 31\,\% (Sch\"uhle \etal, 2000b). Star observations of $\alpha$  Leonis
before and after the attitude loss provide strong evidence that the responsivity
decrease is wavelength dependent with a tendency to become rather small at the
longest wavelengths (Lemaire, 2002). Consequently, we adopted a relative loss in
the responsivities of 31\,\% for wavelengths shorter than 123.8~nm (N\,{\sc v})
(as before), which linearly decreases to 5\,\% at 161~nm.
\end{enumerate}

\begin{figure}[th]
\includegraphics[width=14cm]{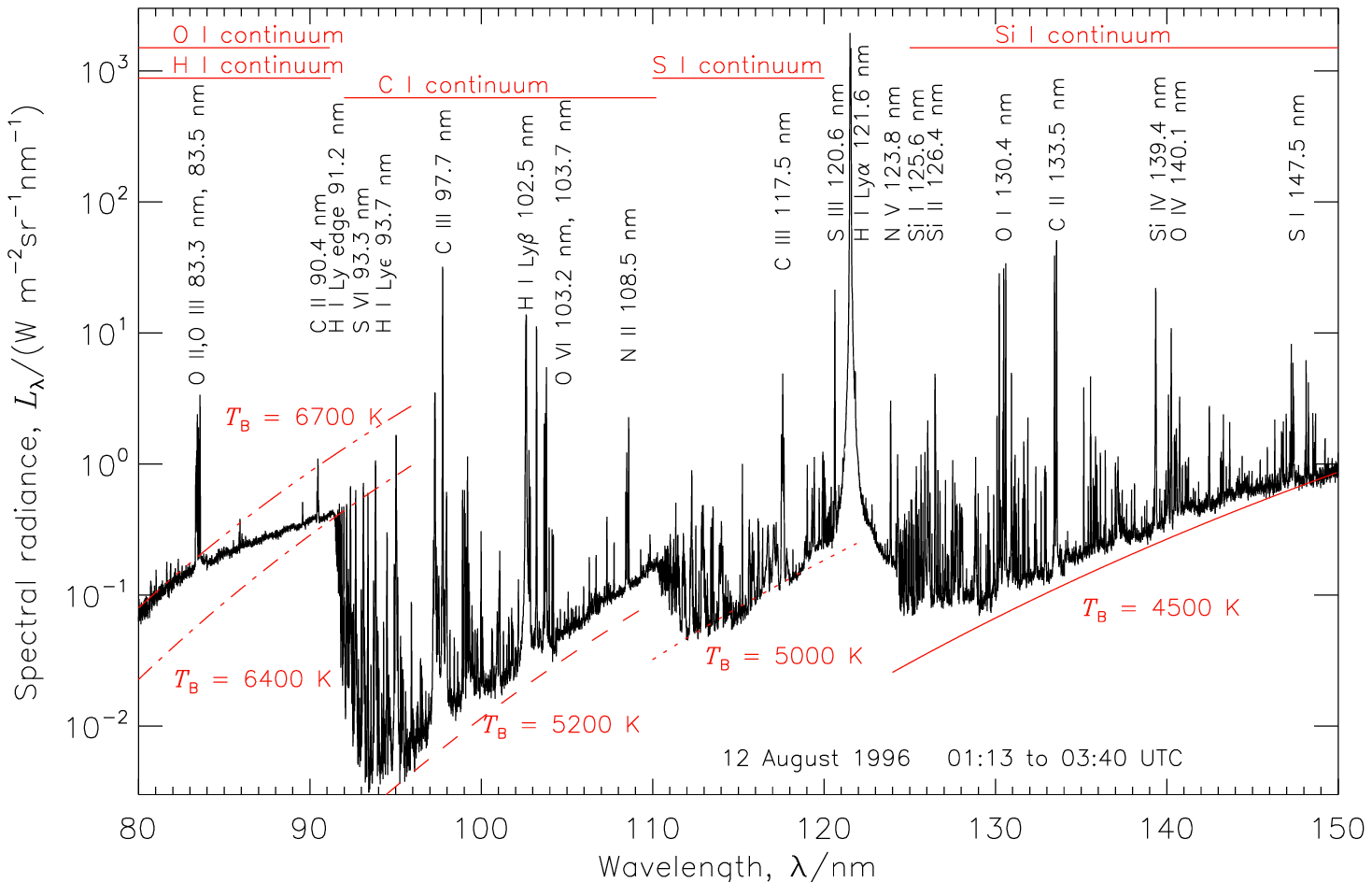}
\caption{\small
Spectral radiance of the quiet Sun in the
VUV range from a region near the centre of the disk.
Prominent emission lines are marked. The spectral radiances expected for some
brightness temperatures, $T_{\rm B}$, are shown in red as approximations of
the continua in the corresponding wavelength ranges
(after Wilhelm and Fr\"ohlich 2010).}
\label{fig:spec}
\end{figure}

The responsivities of SUMER are shown in Figure~\ref{fig:cal} as a typical
result of the ground and in-flight calibration activities.
Note that the radiant energy is measured here as the number of photons with
energy $h\,\nu = h\,c_0/\lambda$, where $h$ is Planck's constant and
$c_0$ the speed of light in vacuum. This convention is often adopted in radiometry.
The spectral responsivity curves displayed in Figure~\ref{fig:cal} refer to
situations with low count rates both for the total detector and for single
pixels. Whenever the total count rate exceeds about $5 \times 10^4$~s$^{-1}$,
a deadtime correction is required; and with a single-pixel rate above
about 3~s$^{-1}$, a gain-depression correction is called for (see Section~3.4).
The uncertainties given in panels (c) and (d) include the contributions of optical
stops and diffraction effects, but the uncertainty of the pixel size has to be treated separately.
When applying spectral responsivity curves as shown in Figure~\ref{fig:cal}(b),
(c) and (d) to the telemetry data, it is necessary to take into account the
effects of field stops as well as the epoch of the observation. This can be
accomplished by applying the SUMER calibration programme {\bf radiometry.pro}
\footnote{\tt sohowww.nascom.nasa.gov/solarsoft/soho/sumer/idl/contrib/wilhelm/rad/}.
The programme can perform all calculations in photon units or in energy
units in accordance with SI. As an example, Figure~\ref{fig:spec} depicts the
VUV radiance spectrum of a QS region with many emission lines and some
continua in the wavelength range from 80~nm to 150~nm.

The responsivity curves are available for the periods from
January 1996 to June 1998 and from November 1998 to December 2001.
They have undergone modifications in the past and might do so again in the
future. There are two reasons for such modifications: an improved
understanding of the performance of the instrument including its calibration;
and changes of the status of the instrument with time.
There are indications that the responsivity of the instrument did not
change at least until April 2007, although the uncertainties increased.
After April 2010, however, a thermal runaway effect in the MCP of detector~B
forced us to reduce the high voltage. This resulted in a significant drop of
the sensitivity and a partial loss of the KBr coated section of the detector.
Since April 2012, only the bare sections of detector~B can be used. Figure~10.2 of
Wilhelm \etal (2002) summarizes the modifications since October 1999 continuing
the history documented in Figure~4 of Wilhelm \etal (2000), demonstrating
that the calibration status in the central wavelength range is very
consistent over time for both detectors. We recommend in general the
joint evaluation, but near 80~nm both detectors cannot be treated jointly.
Thus the separate responsivity of detector~B should be used here.
Before an adequate low-gain level of detector~B was found, a test
configuration was used between 24 September and 6 October 1996.

\subsection{Other Data Reduction Aspects}

Various other data correction algorithms have been established that are used
for special applications, but have not been included in the set of standard
data reduction procedures applied to the data in the archive.

\subsubsection{Wavelength Calibration, Line Identification, Doppler Velocities}
\label{sec: Doppler}

The wavelength setting is accomplished by the linear movement of a rod that changes
the reflection angle of the plane mirror and thus the angle of incidence on the
grating ({\it cf.}, Section~2.2.3).
The relationship between actuator step and wavelength setting is highly
non-linear and has been approximated by a lookup table. This lookup table and
the dispersion relation as discussed before have been used to convert the
spectral pixels to physical units. The wavelength of L1 data is given in unit of
nanometer. The limited accuracy and reproducibility of the wavelength setting
introduce an uncertainty of several pixels, which is far below the spectral resolution.
A more careful wavelength calibration is required, if it is important to know
the absolute wavelength of a spectral line. Since SUMER has no on-board calibration lamp,
each set of spectra with the same wavelength setting has in this case
to be calibrated using nearby photospheric and chromospheric lines of the solar spectrum
as wavelength standards. This cannot be completed in an automated manner and
will be the task of the user. The wavelength calibration in the archive
is based on the nominal dispersion. Moreover, it assumes that the line of
interest is at the central pixel of the spectral window. Unpredictable
offsets of many pixels render this assumption as unrealistic. Therefore, the automated
wavelength scale given in the archive is only a first guess.

Even very faint lines could be detected in deeply exposed on-disk and off-disk
spectra because of the extremely low level of dark counts, and many new line
identifications were possible. A comprehensive overview of the solar spectrum
in the SUMER spectral range is provided in spectral atlases of disk
\citep{Curdt01} and coronal features \citep{Curdt04}.

Centroiding allows to determine the position of unblended spectral lines down
to one tenth of a pixel, in particular for lines observed in second order of diffraction.
The line shift $\delta \lambda$ can be used to calculate the Doppler flow $v_{\rm D}$ using
\begin{equation}
v_{\rm D} = \frac{\delta \lambda}{\lambda}\,c_0
\end{equation}
where $c_0$ is the speed of light in vacuum and $\lambda$ the wavelength of the spectral
line. Several conditions must be met to reach uncertainties as low as
1~km\,s$^{-1}$ to 2~km\,s$^{-1}$:
the line of interest must be unblended and gaussian; its laboratory wavelength must be
well-known; and wavelength standards have to be at rest and close to the line of interest.

The limiting parameter for the effective spectral resolution of the instrument
is certainly given by the detector non-linearity and instability.

\subsubsection{Pixel Shift}
\label{sec: Deltapixel}

The location of the slit image on the detector array is not at all constant.
The size of the slit image varies with wavelength -- an effect of the
wavelength-dependent magnification ({\it cf.}, Section~2.1). The accumulated effect
of the alignment errors between the scan mirror rotation axis, the grating ruling direction,
the direction of the grating focussing mechanism, and the direction of the detector rows
can be evaluated for the distortion-corrected detector arrays (both A and B)
by the function {\bf deltapixel.pro} as shown in Table~\ref{tab:shift}.
The correction for this pixel shift is only needed for co-registration of spectra
with different wavelength settings. For such cases, numerical values can be
extracted from the lookup table in Table~\ref{tab:shift}.
There is no compensation for this effect in the data of the archive.

\begin{table}
\caption{Lookup table. Nominal position of upper and lower pixels of the 120\arcsec~slit at
different wavelengths for detector~B data (for detector~A the wavelengths are
offset by $\approx$12.0 nm to higher values). The shift is given relative to
the slit image at a wavelength setting of 115.2~nm. The shift values are basically
negative which corresponds to an apparent offset towards the North in SUMER coordinates.}
\label{tab:shift}

\begin{tabular}{clllclll}
\hline\noalign{\smallskip}
Wavelength & Lower & Upper & Pixel & Wavelength & Lower & Upper & Pixel \\
$\lambda$ / nm & pixel & pixel & shift &    $\lambda$ / nm & pixel & pixel & shift \\
\noalign{\smallskip}\hline\noalign{\smallskip}
69.2 & 105.6 & 220.9 & -16.8 & 109.2 & 119.5 & 238.7 & -0.9 \\
71.2 & 106.5 & 222.1 & -15.7 & 111.2 & 119.8 & 239.2 & -0.5 \\
73.2 & 107.4 & 223.2 & -14.7 & 113.2 & 119.9 & 239.7 & -0.2 \\
75.2 & 108.2 & 224.1 & -13.8 & 115.2 & 120.0 & 240.0 &  0.0 \\
77.2 & 109.0 & 225.1 & -13.0 & 117.2 & 119.3 & 240.2 &  0.1 \\
79.2 & 109.7 & 226.0 & -12.1 & 119.2 & 119.7 & 240.3 &  0.0 \\
81.2 & 110.5 & 226.8 & -11.3 & 121.2 & 119.4 & 240.2 & -0.2 \\
83.2 & 111.2 & 227.7 & -10.5 & 123.2 & 118.9 & 239.9 & -0.6 \\
85.2 & 112.0 & 228.6  & -9.7 & 125.2 & 118.2 & 239.5 & -1.1 \\
87.2 & 112.7 & 229.5  & -8.9 & 127.2 & 117.4 & 238.9 & -1.9 \\
89.2 & 113.5 & 230.3  & -8.1 & 129.2 & 116.4 & 238.1 & -2.8 \\
91.2 & 114.2 & 231.2  & -7.3 & 131.2 & 115.2 & 237.1 & -3.9 \\
93.2 & 114.9 & 232.1  & -6.5 & 133.2 & 113.8 & 235.9 & -5.1 \\
95.2 & 115.6 & 233.0  & -5.7 & 135.2 & 112.3 & 234.5 & -6.6 \\
97.2 & 116.3 & 233.9  & -4.9 & 137.2 & 110.5 & 232.9 & -8.3 \\
99.2 & 117.0 & 234.8  & -4.1 & 139.2 & 108.6 & 231.1 & -10.1 \\
101.2 & 117.6 & 235.7 & -3.4 & 141.2 & 106.5 & 229.2 & -12.1 \\
103.2 & 118.2 & 236.5 & -2.7 & 143.2 & 104.3 & 227.1 & -14.3 \\
105.2 & 118.7 & 237.3 & -2.0 & 145.2 & 101.8 & 224.9 & -16.7 \\
107.2 & 119.1 & 238.0 & -1.4 & 147.2 &  99.3 & 222.5 & -19.1 \\

\noalign{\smallskip}\hline
\end{tabular}
\end{table}

\subsubsection{Line Broadening}
\label{sec: linewidth}

The width of spectral lines is affected by a contribution of the instrumental
broadening to the Doppler broadening.
Using the function {\bf con-width-funct-3.pro} the instrumental width can be taken
out by using a de-convolution matrix.

\subsubsection{Straylight and Dark Signal}
\label{sec: Straylight}

Because of the excellent surface quality of the primary mirror ({\it cf.},
Section~2.1; `Optical design'), the scattered light of the disk falls off by
five orders of magnitude within 20~arcsec ({\it cf.}, Figure~1
in Lemaire \etal, 1998). It is, therefore, possible to observe the lower
corona off-disk without occultation. At larger limb distances, however, the
off-limb scattered light dominates the spectra. The fall-off curves for the
scattered light levels are the result of the large-angle scatter characteristic
of the micro-roughness of the mirror coating, which is wavelength dependent and also depends
on the non-uniform and variable brightness distribution of the disk. It is therefore not
possible to apply an easy algorithm for automatic straylight subtraction. For
many applications, the scattered light is unproblematic, it may even help with the
wavelength calibration. In all other cases the straylight has to be subtracted by the user.

During the first years in orbit, the data from the photon counting detection system
was practically noise-free ({\it cf.}, Section~2.3 or Wilhem \etal, 1997a).
Only during rare solar energetic particle (SEP) events
with a strong high energy contribution a temporary increase of dark counts was observed.
The rate of dark counts increased, however over the years, and recently, the
number of flaring pixels may become a problem for long exposures at low
signal.

\subsubsection{Thermo-Elastic Deformation Effects}

An analysis by \citet{Rybak99} revealed a parasitic effect of the algorithm
used to regulate the temperature of the optical bench. It was found that the
oscillation of the heater duty cycle had an effect of the line position on
the detector due to thermoelastic deformations.
For the front section of the instrument, \citet{Rybak99} report
oscillation amplitudes of $\approx$0.3~K with a period of $\approx$120~min.
Similarly, in the rear section of the instrument the temperature oscillates with
an amplitude of $\approx$0.1~K and with a period of $\approx$75~min.
As a consequence of these temperature variations, the position of a spectral line
was periodically shifted by up to 2.5~nm, dominated by the front bench.
\citet{Rybak99} also report a correction procedure needed for long-duration
studies during the first years that are sensitive to this effect.

With increasing equilibrium bulk temperatures (ageing effect of the thermo-
optical properties), the need of the heaters was reduced and the heater duty cycle
diminished. Already in 1998, the thermoelastic deformations
became very small and disappeared later.

\section{Archive Description}
\label{sec: Archive}

\begin{figure}[h]
\centering
\includegraphics[width=12cm]{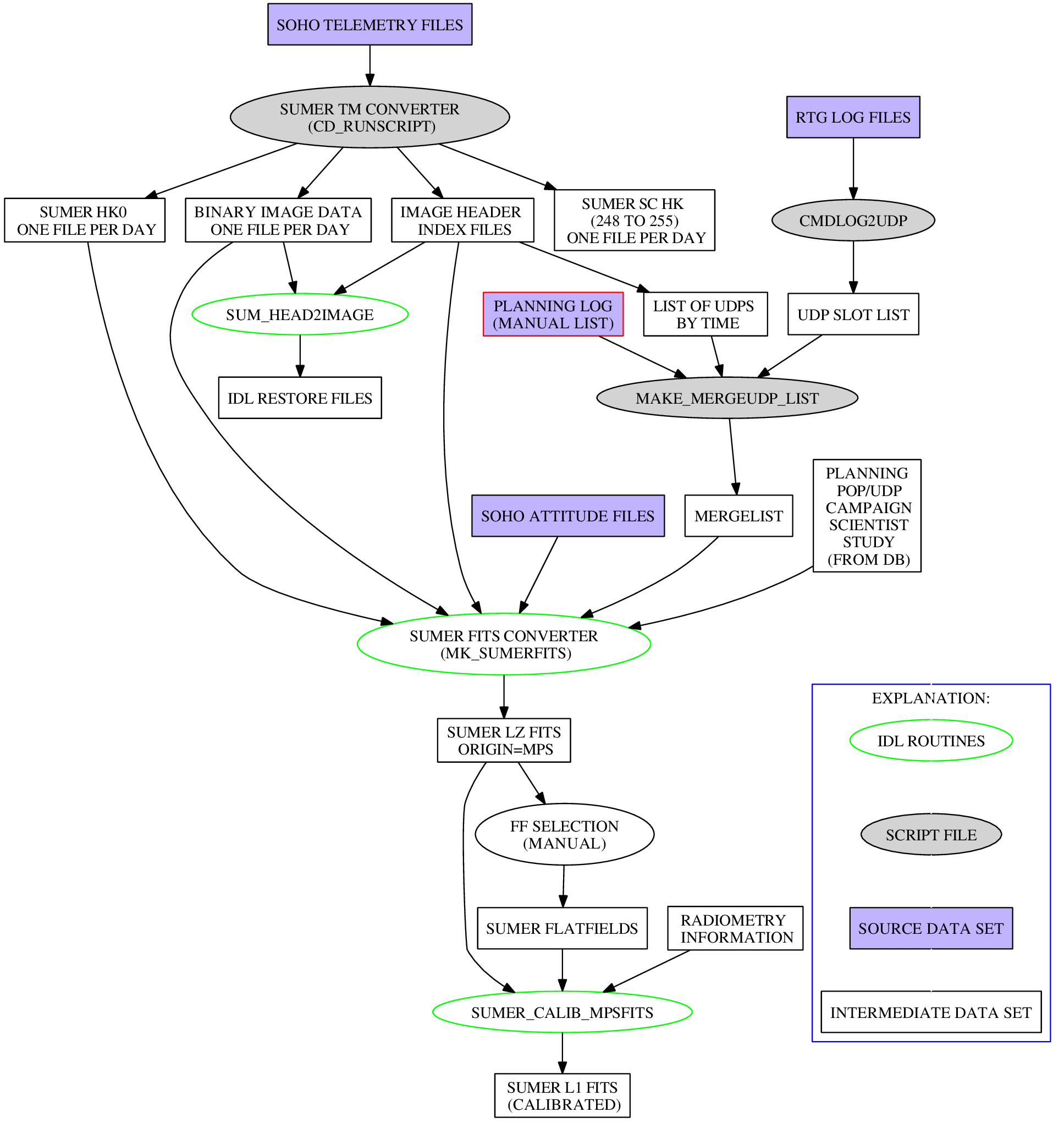}

\caption{Data Flow Diagram.}
\label{fig:dataflow}
\end{figure}

Figure \ref{fig:dataflow} shows the path from the various data sources towards the
SUMER FITS data products, including intermediate processing steps.


\subsection{FITS LZ}
\label{sec: fitslz}
The IDL routine {\bf mk\_sumerfits} takes the binary data from the TM processing
(see Section~\ref{sec: TM processing}) and produces SUMER LZ FITS data.
The main task is to create a FITS header with all the information needed.
In addition some missing HK0 data is correlated to the image (see Section~\ref{sec: HK}).
This routine takes care of assembling several wavelength windows taken with one exposure
into one detector image (see Table~\ref{tab:formats}). This step is necessary because
of the difficulty that further processing steps like geometrical correction
`deform' the image so that a composition of two adjacent images will not be possible without gaps.

The SUMER FITS LZ product is decompressed, solar coordinate corrected (North up) and
wavelength direction corrected (low to high from left to right) data.
All further processing steps described in Section~\ref{sec: Intro} are performed in
the FITS L1 processing which is described in Section~\ref{sec: fitsl1}.

\subsubsection{LZ FITS File Structure}
\label{sec: fitslzfile}
The created SUMER LZ FITS file contains a {\bf{H}}eader {\bf{D}}ata {\bf{U}}nit - the header, and
a {\bf{P}}rimary {\bf{D}}ata {\bf{U}}nit - the image data. In addition there is an extension.

\begin{itemize}
\item[{\bf HDU:}] This is the primary FITS header (an example is given in the
electronic supplementary material).
\item[{\bf PDU:}] This is the actual image data set.
\item[{\bf Extension SUMER-RAW-IMAGE-HEADER:}] The original telemetry image header(s) of the image.
This is a byte array of at least 120 bytes which can be analysed with the SUMER
header routines (see the SUMER Data Cookbook for details). This data is kept for
compatibility so that the `old' routines can also be used to analyse the data.
For each image acquired during the detector integration (1 to 8), there is one 120 byte array.

\end{itemize}

\subsection{FITS L1 - Calibrated Data}
\label{sec: fitsl1}

\subsubsection{Definition}
The SUMER FITS L1 data is defined calibrated data, where all processing steps
described in Section~\ref{sec: Intro} are performed including radiometric calibration.

\subsubsection{Restrictions}
The calibration of SUMER data is only possible for data compressed with the
SUMER lossless compression schemes 6 and below (see SUMER Operations Guide\footnote{
\tt www.mps.mpg.de/projects/soho/sumer/text/sum\_opguide.html}). Images compressed
with other schemes and from the rear slit camera are ignored by the L1 processing.

\subsubsection{Flatfields}
\label{sec: fitsflatfield}
Before starting to describe the production of the calibrated SUMER FITS data (L1) a short description
is given for the preparation of the flatfield data out of the newly created SUMER FITS LZ data.

During the processing of SUMER  LZ data images matching the conditions for SUMER flatfields
are logged with their filename in a list. This list is afterwards taken to produce the flatfield data
for the L1 production process. The IDL routine {\bf sum\_make\_fits\_ff} takes
care of all the necessary steps. One of these steps is to remove the odd-even
pattern from the raw image (see Section~\ref{sec: FF}).

Another step results from the changed acquisition strategy in later times of
SUMER operation. In the beginning a flatfield was taken as one long exposure
and then processed and stored for on-board flatfielding. These data were
downlinked as two separate images, one was the raw data and one the processed on-board flatfield.
Later, due to the `odd-even pattern' (see Section~\ref{sec: FF}, the on-board processing
was skipped and only the raw data were downlinked. To exclude
transmission errors of the flatfield, the flatfield was taken in
several single exposures which were then added to one flatfield image.
This addition is also done by {\bf sum\_make\_fits\_ff}.

In addition this routine `corrects' some image rows which were missing due
to telemetry gaps. The replacement is indicated in the FITS header, as a comment.

\subsubsection{Processing}
\label{sec: L1Processing}

The L1 processing takes LZ data sets and performs all necessary processing to
get the calibrated data. The overall routine {\bf proc\_sumer\_calib} takes care
of the in and outfile organization {\it e.g.} reading LZ files and putting all processed
data into a new file.

All the calibration steps are included in the routine
{\bf sumer\_calib\_mpsfits} which is called by {\bf proc\_sumer\_calib}.
This routine calls the various SUMER calibration routines
in the correct order and logs the performed processing in
the FITS header.

Via parameters, the processing level can be controlled to do
a step-by-step calibration for verification.

\paragraph{Reverse Flatfield}
If an on-board faltfield processing has already been done, this is
reversed for preparation of the odd-even correction.
\newline
The FITS keywords reflecting this processing are:

\vspace{3mm}

\begin{tabular}{l l}
Keyword&Description\\ \hline
SSFF& Mark as not processed on board\\
RFLATFIL&Used reverse flatfield\\
history&Applied {\bf sum\_flatfield} (reverse) \\
\end{tabular}

\paragraph{Dead Time Correction}
The dead time correction step is performed by the {\bf deadtime\_corr}
routine.
\newline
The FITS keywords reflecting this processing are:

\vspace{3mm}

\begin{tabular}{l l}
Keyword&Description\\ \hline
DEADCORR&Dead time correction\\
DCXDLEV&Dead time corr XDL input value\\
history&Applied {\bf deadtime\_corr}\\
\end{tabular}

\paragraph{Odd Even Correction}
The odd even correction is performed by the {\bf sum\_flatfield} giving the
oddeven array for the specified detector as a parameter.
\newline
The FITS keywords reflecting this processing are:

\vspace{3mm}

\begin{tabular}{l l}
Keyword&Description\\ \hline
ODEVCORR&Odd-even correction\\
history&Applied {\bf sum\_flatfield} with odd-even corr\\
\end{tabular}

\paragraph{Local Gain Correction}
Local gain correction is performed by calling the subroutine {\bf local\_gain\_corr}.
\newline
The FITS keywords reflecting this processing are:

\vspace{3mm}

\begin{tabular}{l l}
Keyword&Description\\ \hline
LGAINCOR&Local gain correction\\
history&Applied {\bf local\_gain\_corr}\\
\end{tabular}

\paragraph{Flatfield Correction}

The flatfield used is the one closest in time ahead of the image acquisition date.
The processing itself is done by the function {\bf sum\_flatfield} using the image
and the flatfield data as parameters.
\newline
The FITS keywords reflecting this processing are:

\vspace{3mm}

\begin{tabular}{l l}
Keyword&Description\\ \hline
FLATCORR&Record the processing\\
FLATFILE&Used flatfield data\\
history&Applied {\bf sum\_flatfield}\\
\end{tabular}

\paragraph{Distortion Correction}
Distortion correction for the data is done by calling the function {\bf destr\_bilin}
with the image data as parameter.
\newline
The FITS keywords reflecting this processing are:

\vspace{3mm}

\begin{tabular}{l l}
Keyword&Description\\ \hline
GEOMCORR&Record the processing\\
history&Applied {\bf destr\_bilin}\\
\end{tabular}

\paragraph{Radiometric Calibration}
The radiometric calibration is performed by calling the {\bf s\_fitsrad} subroutine.
This routine takes care of the calculation of the different radiometries
(first or second order, KBr or bare) by calling the
{\bf radiometry} function with the appropriate parameters. Once the radiometry values
for first and second order are calculated, the first order radiometry is applied
to the image data. Both radiometry arrays are then stored in the
FITS file as an extension.
\newline
The FITS keywords reflecting this processing are:

\vspace{3mm}

\begin{tabular}{l l}
Keyword&Description\\ \hline
RADCORR&Radiometry calibration performed\\
RADORDER&Radiometry for wavelength order (first or second order)\\
AVARADO&Available radiometry orders\\
IMGUNITS&Units for image (W sr$^{-1}$ m$^{-2}$ {\AA}$^{-1}$) \\
LEVEL& 1\\
PRODLVL&L1\\
history&Applied {\bf radiometry}\\
\end{tabular}

\paragraph{L1 FITS File Structure}

After having done all the calibration steps and gathered all needed data
the data is written back to the FITS file in the following order:
\begin{itemize}
\item[{\bf HDU:}] This is the primary FITS header (an example is given in the
electronic supplementary material).
\item[{\bf PDU:}] This is the actual calibrated image data set.
\item[{\bf Extension SUMER-RAW-IMAGE-HEADER:}] ({\it cf.}, Section~\ref{sec: fitslzfile}
\item[{\bf Extension SUMER-RAD-ARRAY:}] This extension includes the radiometry arrays for first
and second order. Since only one radiometric calibration can be performed on the data at
a time, the purpose of this data arrays is to be able to reverse the actual radiometric
calibration (indicated in keyword RADORDER) and perform the other one - if available.
The number of available orders/arrays is indicated in the keyword AVARADO.
\end{itemize}

\subsection{Processing Remarks}
The IDL routines used for the FITS creation and calibration processing check on various
FITS keywords for processing conditions and if processing can be performed on
the specified data set. The routines also record performed processing steps in
FITS keywords, so a processing step can not accidently be performed twice on the
data.

\section{Conclusion}
Scientists all over the world who have been using SUMER data have, sharing their
experience, contributed to this work ({\it cf.}, Figure~\ref{fig:publ}).
The benefits of this long
learning process are comprised in the SUMER archive. Here, we have made an effort to
describe the non-trivial task of processing these data in great detail and in a
transparent manner. The trend in Figure~\ref{fig:publ} clearly indicates that there is
still interest in SUMER data and that it is realistic to assume that future work will
come. It is our intention to encourage future users to take advantage of such
ready-to-use data that is not dependant on computer systems. SUMER is now close
to end of the operational life time. Therefore, in the future the archive
will -- with the exception of {\it Interface Region Imaging Spectrograph}
(IRIS; dePontieu \etal., 2012) spectra -- be the only source of data in the SUMER spectral range.
We tried hard to complete the archive so that it can be used for joint science with IRIS.
And we hope that in this new format enough meta information is provided that can be used for data mining.

\begin{sloppypar}
\begin{acknowledgements}
The SUMER project is financially supported by DLR, CNES, NASA, and the ESA PRODEX
Programme (Swiss contribution). SUMER is part of SOHO of ESA and NASA. The
instrument was jointly operated by teams from IAS and MPS. We specially thank
Gilles Poulleau servicing the ground equipment for so many years.
Numerous scientists of the community helped to coordinate the science operations.
\end{acknowledgements}
\end{sloppypar}

{}

\end{article}
\end{document}